\documentstyle[prb,aps,eqsecnum]{revtex}

\begin{document}
\draft
\title{Cumulant expansion of the periodic Anderson model in infinite dimension }
\author{M. E. Foglio\thanks{%
Associate Member of ICTP,Trieste,Italy}}
\address{Instituto de Fisica ``Gleb Wataghin''\\
Universidade Estadual de Campinas,UNICAMP\\
13083-970 Campinas, S\~{a}o Paulo, Brasil}
\author{M. S. Figueira}
\address{Instituto de Fisica, Universidade Federal Fluminense,UFF\\
24000-970 Niter\'{o}i, Rio de Janeiro, Brasil}
\date{\today}
\maketitle

\begin{abstract}
The diagrammatic cumulant expansion for the periodic Anderson model with
infinite Coulomb repulsion ($U=\infty $) is considered here for an
hypercubic lattice of infinite dimension ($d=\infty $). The nearest neighbor
hopping of the uncorrelated electrons is described exactly by a conduction
band, while two different models of hybridization are treated as a
perturbation. The same type of simplifications obtained by Metzner for the
cumulant expansion of the Hubbard model in the limit of $d=\infty $, are
shown to be also valid for the periodic Anderson model. The derivation of
these properties had to be modified because of the exact treatment of the
conduction band. 
\end{abstract}


\section{INTRODUCTION}

The periodic Anderson model (PAM) gives a schematic description of very
important systems with strongly correlated electrons, and there are several
recent reviews on this subject \cite{Reviews}. The model consists of a
lattice with two localized electronic states at each site, strongly
correlated by a Coulomb repulsion $U$, plus a band of uncorrelated
conduction electrons (c-electrons) that hybridize with the localized
electrons (f-electrons). The cumulant expansion of the PAM discussed in this
work is a perturbative expansion around the atomic limit, and is an extension%
\cite{FFM1} of the method originally employed by Hubbard \cite{Hubbard5} to
study his well known model of correlated itinerant electrons\cite{Hubbard123}%
. He introduced operators $X_{j,ba}=\mid j,b\rangle \langle j,a\mid ,$ that
transform the local state $\mid a\rangle $ at site $j$ into the local state $%
\mid b\rangle $ at the same site, and developed a diagrammatic method that
circumvents the fact that the $X$ operators are neither fermions nor bosons.
His work was the first application of the cumulant expansion to a quantum
system of fermions\cite{Wortis74} and has several desirable properties: it
seems to be a natural extension of the usual diagrammatic techniques, it
does not have excluded site problems in the lattice summations and it is
possible to derive a linked cluster expansion for the grand canonical
potential( see ref. \cite{FFM1} for more details). At the same time, the use
of $X$ operators allows to work in a subspace of the whole space of local
states, that contains only the states that are relevant to the problem of
interest. It seems therefore useful to understand better this method, and
the main purpose of the present paper is to study the properties of the
cumulant expansion for the PAM in the limit of infinite spatial dimensions ($%
d\rightarrow \infty $).

Metzner and Vollhardt\cite{MetznerVoll89} have recently applied the limit $%
d\rightarrow \infty $ to strongly correlated fermion systems, providing
non-trivial models of the Hubbard type that are substantially simpler to
analyze. This development has stimulated many new works, and we could
mentioned the very successful Dynamical Mean Field Theory\cite{GeorgesKKR},
that uses the local properties of the self energy for $d\rightarrow \infty $
as a starting point. Several methods have been employed to study the PAM in
this limit\cite{Jarrel,MutouH1,MutouH2,SasoI}, and the cumulant expansion of
the Hubbard model was reconsidered by Metzner\cite{Metzner91}, showing that
there is an important reduction in the number and type of cumulant diagrams
that appear in the expansion when $d\rightarrow \infty $ in an hypercubic
lattice. In the Hubbard model there is no hybridization and the hopping
Hamiltonian is used as perturbation, while the PAM is usually employed to
model systems with a bandwidth much larger than the hybridization energies,
and it is then preferable to diagonalize the conduction band and use the
hybridization Hamiltonian as perturbation. It was then necessary to make
substantial modifications to Metzner's derivation in order to apply it to
the PAM when $d\rightarrow \infty $.

In the following sections two types of hybridization models are considered:
the purely ``local'' hybridization and the ``nearest neighbor'' (n.n.) one.
In the n.n. hybridization there are electronic transitions between localized
(``f'') states at a site and conduction Wannier states at the nearest
neighbor sites, while in the local hybridization those transitions only
occur at the same site. The same type of diagram cancellation obtained by
Metzner for the Hubbard model is obtained for these two hybridization models
in the PAM.

As the local hybridization between $f$ and $d$ or $s$ electrons should
vanish when the site has inversion symmetry, the n.n. hybridization
discussed in this paper is more realistic than the local one, because it
does not necessarily vanish for that type of site. The local approximation
is usually employed because it makes the calculation simpler; and the
changes introduced by an interatomic hybridization have been already
discussed in the case of a diatomic-molecule model\cite{LeoChino}.

\section{The hypercubic lattice}

\label{Sec2} Let us consider free electrons in an hypercubic lattice of
dimension ${\it d} $, described by a Hamiltonian\cite{MHartmann89}

\begin{equation}
H_c=\sum_{{\bf n},{\bf m,\sigma }}t_{{\bf n},{\bf m}}\ C_{{\bf n,\sigma }%
}^{\dagger }\ C_{{\bf m,\sigma }}  \label{E2.1}
\end{equation}
\noindent with translational invariance. The position of the site ${\bf n}%
=(n_1,\ldots ,n_d)$ is given by ${\bf R}_{{\bf n}}=\sum_{\alpha =1}^d{\bf e}%
_\alpha \ n_\alpha $, where $\left| {\bf e}_\alpha \right| =a$ is the
lattice parameter and the $n_j$ are integers. Only n.n. hopping shall be
considered, so that $t_{{\bf n},{\bf m}}$ is non-zero and equal to $-%
\overline{t}$ when the components $m_{\alpha \text{ }}$ of ${\bf m}$ satisfy
$m_\alpha =n_\alpha +\delta _{\alpha ,\gamma }$ for $\gamma =1,2,\ldots ,d$.
Employing the Wannier transformation

\begin{equation}
C_{{\bf n,\sigma }}=\frac{1}{\sqrt{N}}\sum_{{\bf k{,\sigma }}}\ C_{{\bf k{%
,\sigma }}}\ \exp [i{\bf k}.{\bf R}_{{\bf n}}]\quad ,  \label{E2.2}
\end{equation}
\noindent with ${\bf k}$ satisfying cyclic boundary conditions, one obtains

\begin{equation}
H_c=\sum_{{\bf k,\sigma }}E_{{\bf k}}\ C_{{\bf k,\sigma }}^{\dagger }\ C_{%
{\bf k,\sigma }}\quad ,  \label{E2.3}
\end{equation}
\noindent where

\begin{equation}
E_{{\bf k}}=-2\ t\sum_{s=1}^d\cos (k_s\ a)\quad .  \label{E2.4}
\end{equation}

The electronic Green's functions (GF) for imaginary time are defined by\cite
{FFM1}
\begin{equation}
G_{{\sigma }}({\bf k,}\tau )\equiv \left\langle \left( C_{{\bf k,\sigma }%
}(\tau )\ C_{{\bf k,\sigma }}^{\dagger }\right) _{+}\right\rangle \quad ,
\label{E2.5}
\end{equation}
\noindent where $C_{{\bf k,\sigma }}(\tau )\equiv \exp [H_{c}\tau ]\ C_{{\bf %
k,\sigma }}\ \exp [-H_{c}\tau ]$ and the subindex $+$ in Eq.(\ref{E2.5}) is
the usual chronological ordering of Fermi operators, with $\tau $ increasing
to the left. To emphasize that the hybridization was not considered in these
GF, they will be denoted with $G_{{\sigma }}^{0}({\bf k,}\tau )$, and after
Fourier transformation in the imaginary time $\tau $ they are given by\cite
{FF1} $G_{{\sigma }}^{0}({\bf k,}\omega _{\nu })=-(i\omega _{\nu }-E_{{\bf k}%
})^{-1}$. To obtain a finite average energy per site when $d\rightarrow
\infty $, it is necessary to renormalize $H_{c}$, taking a finite $t$ in the
non-zero $t_{{\bf n},{\bf m}}=-t/\sqrt{2d}$ (i.e. when ${\bf n}$ and ${\bf m}
$ are n.n.). In that limit the energy density of states is given by\cite
{MHartmann89}

\begin{equation}
\rho _d(E)=\frac 1{\sqrt{2\pi t^2}}\exp \left[ -\frac{E^2}{2\ t^2}\right]
\quad ,  \label{E2.6}
\end{equation}
\noindent and

\begin{equation}
\left\langle \left( E({\bf k})\right) ^2\right\rangle =t^2\quad .
\label{E2.7}
\end{equation}

The exact solution of $H_{c}$ is used in the present treatment of the PAM,
employing the hybridization as perturbation. To study the behavior of the
cumulant expansion in the limit $d\rightarrow \infty $, it is convenient to
consider the direct space GF, given by

\begin{equation}
G_{{\sigma }}^0({\bf R}_{{\bf n},{\bf m}},\omega _\nu ){\sl \equiv }\frac 1N%
\sum_{{\bf k}}G_{{\sigma }}^0({\bf k,}\omega _\nu )\exp [i{\bf k}.{\bf R}_{%
{\bf n},{\bf m}}]\quad ,  \label{E2.8}
\end{equation}
where ${\bf R}_{{\bf n},{\bf m}}={\bf R}_{{\bf n}}-{\bf R}_{{\bf m}}$, as
well as its dependence with the minimum number $p$ of n.n. jumps necessary
to go from ${\bf m}$ to ${\bf n}$. When the GF are obtained for large $d$
using the hopping as a perturbation, one immediately shows that $G_{{\sigma }%
}^0({\bf R}_{{\bf n},{\bf m}},\omega _\nu )\simeq O\left| \theta ^p\right| $%
,where $\theta =t/\sqrt{2d\text{ }}$. A similar result is obtained for the
``exact'' $G_{{\sigma }}^0({\bf R}_{{\bf n},{\bf m}},\omega _\nu )$ of Eq. (%
\ref{E2.8}), but the discussion will be postponed to a latter section, in
which a more general property will be shown.

\section{THE HAMILTONIAN OF THE PAM}

\label{Sec3} The Hamiltonian of the whole system is

\begin{equation}
H=H_c+H_f+H_{h\qquad ,}  \label{E3.1}
\end{equation}
\noindent where $H_c$ is the Hamiltonian of the conduction electrons
(c-electrons) discussed in Section \ref{Sec2}. The second term $%
H_f=\sum_{j,\sigma }E_\sigma X_{j,\sigma \sigma }$ describes independent
localized electrons (f-electrons), where a simple index $j$ has been used to
indicate the sites ${\bf n}$. The last term is the hybridization Hamiltonian%
\cite{FFM1}, giving the interaction between the c-electrons and the
f-electrons:

\begin{equation}
H_h=\sum_{j,{\bf k},\sigma }\left( V_{j,{\bf k},\sigma }\ X_{j,0\sigma
}^{\dagger }\ C_{{\bf k},\sigma }+V_{j,{\bf k},\sigma }^{*}\ C_{{\bf k}%
,\sigma }^{\dagger }\ X_{j,0\sigma }\right) \quad .  \label{E3.2}
\end{equation}

The state space of the f-electrons, at a given site j, is spanned by four
states: the vacuum state $\mid j,0\rangle ,$ the two states$\mid j,\sigma
\rangle $ of one f-electron with spin component $\sigma $ and the state $%
\mid j,2\rangle $ with two electrons of opposite spin. In the limit of
infinite electronic repulsion ($U=\infty $) considered here, the state $\mid
j,2\rangle $ is always empty, and can be projected out of the space. The
operator $X_{j,0\sigma }$ destroys the electron in state $\mid j,\sigma
\rangle $ leaving the site in the vacuum state $\mid j,0\rangle $ with
energy $E_{j,0}=0$, and its Hermitian conjugate $X_{j,0\sigma }^{\dagger
}=X_{j,\sigma 0}$ reverses that process. The Hubbard operators $X_{j,ab}$
are not usually Fermi or Bose operators, and a product rule:

\begin{equation}
X_{j,ab}\ X_{j,cd}=\delta _{bc}\ X_{j,ad}\qquad ,  \label{E3.3}
\end{equation}
\noindent should be employed when the two operators are at the same site.
For different sites one should first classify the operators in two families:
$X_{j,ab}$ is of the ``Fermi type'' when $\mid a\rangle $ and $\mid b\rangle
$ differ by an odd number of Fermions, and it is of the ``Bose type'' when
they differ by an even number of Fermions. At different sites, two
X-operators of the Fermi type anticommute, and they commute when at least
one of them is of the Bose type. To complete the definition, the $X_{j,ab}$
should anticommute (commute) with the $C_{{\bf {k}},\sigma }^{\dagger }$ and
$C_{{\bf {k}},\sigma }$ when they are of the Fermi type (Bose type).
Sometimes it will be convenient to employ $X(\gamma )$ to describe the
Hubbard operators, where $\gamma =(j,\alpha ,u)$, and $\alpha =(b,a)$ gives
the transition $\mid a\rangle \rightarrow \mid b\rangle $ that destroys one
electron, i.e. with $\mid b\rangle $ having one electron less than $\mid
a\rangle $ as in Ref. \onlinecite{FF1}, and it is often convenient to
abbreviate $(\gamma ,\tau )=\ell $. The index $u$ is then introduced to
describe the inverse transition , so that $X(\gamma )=X_{j,\alpha }$ when $%
u=-$ and $X(\gamma )=X_{j,\alpha }^{\dagger }$ when $u=+$, and for the PAM
with $U=\infty $ there are only two possible transitions $\alpha =(0,\sigma
) $.

The Grand Canonical Ensemble (GCE) of electrons is employed in this problem,
and it is useful to introduce

\begin{equation}
{\cal H}=H-\mu \left( \sum_{{\bf k},\sigma }C_{{\bf k},\sigma }^{\dagger }C_{%
{\bf k},\sigma }+\sum_{j,a}\nu _{a}\ X_{j,aa}\right) \ ,  \label{E3.4}
\end{equation}
\noindent where $\nu _{a}$ is the number of electrons in state $\mid
a\rangle $ and $\mu $ is the chemical potential. As usual ${\cal H}$ is
split into
\begin{equation}
{\cal H}={\cal H}_{0}{\cal +}H_{h}\ ,  \label{E3.5}
\end{equation}
\noindent where $H_{h}$ will be the perturbation Hamiltonian, and the exact
and unperturbed averages of any operator $A$ will be denoted respectively by
$<A>_{{\cal H}}$ and $<A>$. It is also convenient to denote the exact or
Heisenberg ``$\tau $ evolution'' of any operator $A$ with
\begin{equation}
\widehat{A}\left( \tau \right) =\exp [{\cal H}\tau ]\ A\ \exp [-{\cal H}\tau
]  \label{E3.5a}
\end{equation}
and employ $A\left( \tau \right) =\exp [{\cal H}_{0}\tau ]\ A\ \exp [-{\cal H%
}_{0}\tau ]$ for the unperturbed case. It is convenient to use the
expressions
\begin{equation}
\varepsilon _{j,a}=E_{j,a}-\nu _{a}\mu \qquad \text{and\qquad }\varepsilon _{%
{\bf k,}\sigma }=E_{{\bf k,}\sigma }-\mu \quad ,  \label{E3.8}
\end{equation}
\noindent because the energies appear almost always in those combinations.

The c-electron GF employed in the cumulant expansion are:
\begin{equation}
G_{c{,}\sigma }^{o}({\bf k,}\omega )=-\frac{1}{i\omega -\varepsilon _{{\bf k,%
}\sigma }}\quad ,  \label{E3.9a}
\end{equation}
\noindent where the subindex c was added to differentiate them from the
unperturbed GF of the f-electron.

\begin{equation}  \label{E3.9b}
G_{f,\sigma }^o(\omega )=-\frac{D_\sigma ^0}{i\omega -\varepsilon _\sigma }%
\quad ,
\end{equation}
\noindent where $D_\sigma ^0=\langle X_{\sigma \sigma }\rangle +\langle
X_{00}\rangle $.

\subsubsection{The two Hybridization models}

The general hybridization coefficients are given by (cf. Eq. (2.3) in Ref. %
\onlinecite{FFM1})

\begin{equation}  \label{E3.10}
V_{j,{\bf k,\sigma }}=\frac 1{\sqrt{N_s}}V_\sigma ({\bf k)\exp {%
(ik.R_j)\qquad .}}
\end{equation}

The hybridization is purely local when $V_\sigma ({\bf k)=V_\sigma ^0}$ is
independent of ${\bf k}$, because using the Wannier transformation (cf. Eq.(%
\ref{E2.2})) in Eq. (\ref{E3.2}) it follows that
\begin{equation}
H_h=\sum_{j,\sigma }\left( V_\sigma ^0\ X_{j,0\sigma }^{\dagger }\ C_{{j}%
,\sigma }+(V_\sigma ^0)^{*}\ C_{{j},\sigma }^{\dagger }\ X_{j,0\sigma
}\right) \quad .  \label{E3.11}
\end{equation}

The $H_h$ that corresponds to n.n. hybridization is

\begin{equation}
H_h=\sum_{j,\delta ,\sigma }\left( V_\sigma \ X_{j+\delta ,0\sigma
}^{\dagger }\ C_{{j},\sigma }+(V_\sigma )^{*}\ C_{{j},\sigma }^{\dagger }\
X_{j+\delta ,0\sigma }\right)  \label{E3.12}
\end{equation}
\noindent where the vectors $\delta $ give the position of the $d$ n.n.
sites of the origin, and the corresponding $V_\sigma ({\bf k)}$ is
immediately obtained:
\begin{equation}
V_\sigma ({\bf k)=2\ V_\sigma \sum_{\alpha =1}^d\cos (k_\alpha \ a)=-\frac{%
V_\sigma }t\ E_{k,\sigma }\quad .}  \label{E3.13}
\end{equation}

\subsection{The chain approximation}

In figure~\ref{F1} are shown some of the infinite diagrams that contribute
to the exact GF $\left\langle \left( \widehat{X}_{j,\alpha }(\tau )\
\widehat{X}_{j^{\prime },\alpha ^{\prime }}\right) _{+}\right\rangle _{{\cal %
H}}$. The full circles (f-vertices) correspond to the cumulants of the
f-electrons and the empty ones (c-vertices\ ) to those of the c-electrons.
Each line reaching a vertex is associated to one of the $X$ operators of the
cumulant, and the free lines (i.e. those that do not join an empty circle)
correspond to the external $X$ operators appearing in the exact GF. An
explicit definition of the cumulants can be found in the references \cite
{FFM1,Hubbard5,FF1},and they can be calculated by employing a generalized
Wick's theorem\cite{FF1,Hewson,YangW}.

The first diagram in figure~\ref{F1}a corresponds to the simplest free
propagator $\left\langle \left( X_{j,\alpha }(\tau )\ X_{j^{\prime },\alpha
^{\prime }}\right) _{+}\right\rangle $, and the second diagram in that
figure has an empty circle that corresponds to the conduction electron
cumulant, equal to the free propagator $\left\langle \left( C_{k\sigma
}(\tau )\ C_{k\sigma }^{\dagger }\right) _{+}\right\rangle $. The
interaction is represented by the lines (edges) joining two vertices and,
because of the structure of the hybridization, they always join a c-vertex
to an f-vertex; the number of edges in a diagram gives its order in the
perturbation expansion.

Cumulants containing statistically independent operators are zero, and those
appearing in the present formalism (with the hybridization as perturbation)
vanish unless they contain only $X$ operators at the same site or only $C$
or $C^{\dagger }$ operators with the same $k$ and $\sigma $. The only
non-zero c-cumulants are of second order, because the uncorrelated
c-operators satisfy Wick's theorem. On the other hand, the f-vertices can
have many legs, all corresponding to $X$ operators at the same site, like
the fourth and sixth order cumulants appearing in the rather more
complicated diagram shown in figure~\ref{F1}c.

\label{Sec3A}All the infinite diagrams that contribute to the GF of the
f-electron with cumulants of at most second order are shown in figure~\ref
{F1}a, and they define the ``chain approximation'' (CHA) when all the other
diagrams are neglected. The corresponding approximation for the GF of the
c-electrons corresponds to the diagrams of figure~\ref{F1}b. The diagrams of
the CHA usually appear as part of more complicated diagrams, and it is then
useful to analyze their behaviour when $d\rightarrow \infty $. In the CHA,
the GF is given in frequency and ${\bf {k}}$ space by (cf. Eq. (3.10) in
Ref. \cite{FF1})

\begin{equation}
G_{f,\sigma }({\bf {k},\omega )=\frac{-(i\omega -\varepsilon _{{k}\sigma
})D_\sigma ^0}{(i\omega -\varepsilon _\sigma )(i\omega -\varepsilon _{{k}%
\sigma })-\mid V({k})\mid ^2D_\sigma ^0}\qquad .}  \label{E3.14}
\end{equation}
\noindent and it would be useful to transform it back to real space to show
the dependence with the distance ${\bf {R}}_{{i,j}}$ between the two sites $%
i $ and $j$, as it was done in Eq. (\ref{E2.8}) for the conduction
electrons. Because of the lattice translational invariance, it is enough to
use the distance ${\bf {R}}_j$ of the site $j$ to the origin, and
characterize this site with an ${\bf {n}}=(n_1,\ldots ,n_d),$ (cf. Sec. \ref
{Sec2}). The set of indices $r$ with $n_r\neq 0$ will be denoted with $%
\{r\}_j$, while $s(j)$=$\sum_1^dn_s$ is the minimum number of n.n. jumps
necessary to go from the origin to the site $j$. In both the local and the
n.n. hybridization models, the $G_{f,\sigma }({\bf {k},\omega )}$ depends on
${\bf {k}}$ only through the $\varepsilon _{{\bf {k}}\sigma }=E_{{\bf {k}%
\sigma }}-\mu $, and for any given site $j$ one can write
\begin{equation}
E_{{\bf {k}\sigma }}=E_\sigma (\{r\}_j,{\bf {k})-\theta \sum_{s\in
\{r\}_j}\cos (a\ k_s)\quad ,}  \label{E3.15}
\end{equation}
\noindent where

\begin{equation}
E_\sigma (\{r\}_j,{\bf {k})=-\theta \sum_{s\notin \{r\}_j}\cos (a\ k_s)\quad}
\label{E3.16}
\end{equation}
and $\theta =t/\sqrt{2d}$. Substituting this relation in Eq. (\ref{E3.14})
one can expand it in a power series of all the $\cos (a\ k_s)$ with $s\in
\{r\}_j$ and then transform back to real space to obtain $G_{f,\sigma }({%
{\bf {R}}}_j,\omega ).$ Employing the relation

\begin{equation}
\sum_{k_s}\cos ^m(k_s\ a)\ \exp (in_sk_sa)=0\quad \text{for}\quad
m<n_s\qquad ,  \label{E3.17}
\end{equation}
\noindent which is a consequence of the cyclic boundary conditions, it
follows that

\begin{equation}  \label{E3.18}
G_{f,\sigma }({\bf R}_j,\omega )=O\left| (2d)^{-s(j)/2}\right| \quad \cdot
\end{equation}

The GF for the conduction electrons in the CHA, corresponding to the
diagrams in Fig. \ref{F1}b, is given by:

\begin{equation}
G_{c,\sigma }({\bf {k},\omega )=\frac{-(i\omega -\varepsilon _\sigma )}{%
(i\omega -\varepsilon _\sigma )(i\omega -\varepsilon _{{k}\sigma })-\mid V({k%
})\mid ^2D_\sigma ^0}\qquad ,}  \label{E3.19}
\end{equation}
\noindent and employing the same derivation used above for the f-electrons
it follows that $G_{c,\sigma }({\bf {R}}_j,\omega )=O\left|
(2d)^{-s(j)/2}\right| $. The same relation is obtained in the absence of
hybridization, as stated at the end of Section \ref{Sec2} for $G_{{\sigma }%
}^0({\bf {R}}_{{\bf {n}},{\bf {m}}},\omega _\nu )$.

To close this section let us emphasize that the present expansion employs $%
H_h$ as perturbation, and that the exact solution of the conduction problem
in the absence of hybridization is included in the zeroth order Hamiltonian.
The contribution to the GF joining two sites separated by $s(j)$ n.n. jumps,
that is of order $\left| (2d)^{-s(j)/2}\right| $, includes then
contributions of any order in the $H_c$. It is because of this difference
that the derivation of Metzner\cite{Metzner91} for the Hubbard model had to
be modified for the present problem.

All the contributions of $H_c$ would disappear in the case of a band with
zero width, but the electronic wave functions would still be extended for
the model with n.n. hybridization.

\section{THE CONTRIBUTION OF DIAGRAMS FOR INFINITE DIMENSION}

\label{Sec4}In the present section it will be shown that the only diagrams
that remain in the limit $d\rightarrow \infty $ are those that are
topologically ``fully two particle reducible'' (f.t.p.r.). These diagrams
are defined as those in which any pair of vertices can be separated by
cutting one or two edges\cite{MetznerVoll90}, and are ``topologies
constructed by linking polygons''\cite{Metzner91}. Two points should be made
here: the first is that two different polygons (also called loops or rings
in what follows) can have at most one vertex in common in the f.t.p.r.
topology (see Fig \ref{F2}a,b). The second point, already stressed by
Metzner for the Hubbard model,\cite{Metzner91} is that the topology of a
diagram can be different from those of its possible embeddings in the
lattice, because in the cumulant expansions there is no excluded site
restriction in the lattice sums\cite{FFM1}, and two different vertices of a
diagram can occupy the same site (see Fig. \ref{F2}c,d). The property stated
above refers to the topology of the embeddings, and in the sum of
contributions of diagrams that are not f.t.p.r. there may be terms that give
non zero contributions because they correspond to a f.t.p.r. topology of the
embedding: the diagram in Fig. 2c could contribute when $j=i$, because the
topology of its embedding, given by Fig. 2d, is f.t.p.r.

To prove this property it is necessary to modify the derivation given by
Metzner for the Hubbard model, for the same reasons given in Section \ref
{Sec3A} for the GF dependence on $s(i)$. The proof of this property is given
in the following two subsections, and it can be summarized as follows.
Consider first the diagram in Fig. \ref{F3}a, that corresponds to the sum of
loops with all possible lengths. All the embeddings of these diagrams are
f.t.p.r., and they give a non-zero contribution to the calculation of the
free energy. As a second step, one can consider diagrams like those in Figs.
\ref{F3}c,d, obtained by linking loops of any length at different
f-vertices. One could obtain all these diagrams by a procedure similar to
the vertex renormalization\cite{FFM1,Wortis74,Metzner91}, with the
difference that only insertions of a rather special type are considered. All
these diagrams, as well as their embeddings, are f.t.p.r., and a finite
contribution of all of them is expected. The next step is to consider
diagrams that are not f.t.p.r., like the one shown in Fig \ref{F3}e, that
can be obtained by joining two loops (like that in Fig. \ref{F3}a) at two
different f-vertices. If the family in Fig. \ref{F3}a{\sl \ }gives a finite
contribution, one can show that all the contributions to the diagram in Fig.
\ref{F3}e will vanish when the topology of the embedding coincides with that
of the diagram, i.e. when the embedding is not f.t.p.r., as it would happen
when $i\neq j$. In the special case of $i=j$, (cf. Figs. \ref{F2} c,d for a
special case of this situation) the embedding is f.t.p.r., and the diagram
contribution does not necessarily vanish. Of the two sums over the lattice
sites $i$ and $j$, only one over $i=j$ remains in the limit $d\rightarrow
\infty $, and this simplifies the calculation of this diagram. In
particular, the restriction that in reciprocal space there must be
conservation of ${\bf k}$ at each vertex is removed\cite
{MetznerVoll89,MHartmann89}, leaving only a single conservation of ${\bf k}$
for all the edges joining the collapsed vertices: this makes the calculation
of the diagram much simpler in the limit $d\rightarrow \infty $.

\subsection{The f.t.p.r. diagrams}

\label{Sec4a}

To prove that the contribution of a f.t.p.r. diagram can be finite, let us
consider the family of diagrams represented in Fig. \ref{F3}a. Their
contribution does not vanish and can be expressed as
\begin{equation}
\sum_{j_1}\int d\ell _1\int d\ell _2\ S_2^0(j_1;\ell _1,\ell _2)\
M_2^0(j_1;\ell _1,\ell _2)  \label{E4.1}
\end{equation}
\noindent where
\begin{equation}
M_2^0(j_1;\ell _1,\ell _2)=\left\langle \left( X(\ell _1)X(\ell _2)\right)
_{+}\right\rangle _c  \label{E4.2}
\end{equation}
\noindent is a local cumulant\cite{FFM1} and $\int d\ell _s=\sum_{\alpha
_s}\sum_{u_s}\int_0^\beta d\tau _s$. The abbreviations $s\equiv $ $\ell _s$
and $\int ds\equiv $ $\int d\ell _s$ will be used when there is no
possibility of confusion. The symbol $S_2^0(j_1;\ell _1,\ell _2)\equiv
S_2^0(j_1;1,2)$ corresponds closely to the ``self-field''\cite{Wortis74} $%
S_m(j_1;\ell _1,\ell _2,\ldots ,\ell _m)\equiv S_m(j_1;1,2,\ldots ,m)$, that
was employed in Ref. \onlinecite{FF1} to renormalize vertices, but it gives
only ``insertions'' obtained from simple loops of any length, as represented
by the diagrams in Fig \ref{F3}b; by it definition, all the $\ell _s$
correspond to the same site, indicated by $j_1$. The notation $%
M_2^0(j_1;\ell _1,\ell _2)\equiv M_2^0(j_1;1,2)$ has the same meaning, and
reflects the fact that this cumulant is zero unless $j_1=j_2$ because $%
X(\ell _1)$ and $X(\ell _2)$ are statistically independent in the
unperturbed system when $j_1\neq j_2$. Note that $S_2^0(j_1;\ell _1,\ell _2)$
depends explicitly on the parameters $j_1;\ell _1,\ell _2$ (where $\ell _i$
represents $j_i,u_i,\alpha _i=(0,\sigma _i)$ for $i=1,2$) through the two
hybridization constants $V_{j,{\bf k,\sigma }}$ associated to the edges
joining the insertion vertex.

To consider the addition of a simple loop to any f-vertex of a diagram, it
is convenient to consider first the simplest possible case, shown in Fig.
\ref{F3}c: its contribution is obtained by substituting the cumulant $%
M_2^0(j_1;1,2)$ in Eq. \ref{E4.1} by
\begin{equation}
\int d3\int d4\ M_4^0(j_1;1,2,3,4)\ S_2^0(j_1;3,4)\quad .  \label{E4.3}
\end{equation}
\noindent When the loop is added to a site $j_1$ that is already joined by $%
n $ edges, the corresponding cumulant $M_m^0(j_1;1,2,\ldots ,m)$ suffers a
similar substitution. Repeated application of this procedure at all f
vertices gives all the possible f.t.p.r diagrams, and as the cumulants are
independent of the lattice dimension $d$ and of the site $j$, this procedure
should not affect the order of the contribution with respect to $d$. The
simple loops of Fig. \ref{F3}a give a finite contribution, so that the
contribution of any f.t.p.r diagrams is then of $O\left( \left| d\right|
^0\right) $.

A similar type of procedure can be applied to the CHA\ diagrams of the
one-particle GF, by successive decoration of the f-vertices with any number
of insertions corresponding to the $S_2^0(j_1;\ell _1,\ell _2)$ discussed
above. It is then clear that these diagrams will still be f.t.p.r., and that
the corresponding GF joining two sites separated by $s$ n.n. jumps will be
of order $O\left| (2d)^{-s/2}\right| ,$ as it was shown in Section \ref{Sec3}
for the CHA.

\subsection{The diagrams that are not f.t.p.r.}

\label{Sec4b}

To analyze the diagrams that are not f.t.p.r., consider any one of them as a
``mother'' diagram, and split it into two or more f.t.p.r linked
``daughter''diagrams without any edges in common but such that any of them
has at least two vertices in common with another daughter diagram, as well
as two edges arriving at each of the common vertices. It is clear that any
pair of vertices that are common to two daughter diagrams can not be
separated in the mother diagram by cutting one or two edges, and one says
that they are not ``two particle reducible'' (t.p.r.).

From all the daughter diagrams choose one as a vacuum diagram and transform
the remaining ones into GF by adding external lines of the Bose type (as
indicated in Appendix \ref{ApA}) to all the vertices that each of them had
in common with any other daughter diagram in the mother diagram. Three
examples of this procedure are given in Fig \ref{F4}.

Consider now the real space calculation of the contribution of the daughter
diagrams. In Appendix \ref{ApA} it is shown that by joining one daughter GF
to another diagram and fixing the position of the common vertices, one
obtains the order of the contribution of the resulting diagram as the
product of those of the building diagrams. In Appendix \ref{ApB} it is shown
that unless all the external f-vertices of one daughter GF coincide at the
same lattice site, the corresponding contribution is of $O\left| \left(
\theta \right) ^{p}\right| $ with $p\geq 1$, where $\theta =t/\sqrt{2d}$.
Taking now the only daughter vacuum diagram and adding successively all the
other daughter GF diagrams, the total contribution for given positions of
all the GF external vertices will then be $O\left| \left( \theta \right)
^{p}\right| $ with $p\geq 1$, unless all the external vertices of each
daughter GF coincide at the same lattice site, that can be different for
different daughter GF. A typical example of this type of diagram is given in
Fig \ref{F5}b, while in Fig \ref{F5}c it is shown the corresponding f.t.p.r.
topology of the embedding when the vertices that are not t.p.r. in the
diagram are at the same lattice site{\sl .}

It is then clear that in the sum over the vertices that are common to all
the daughter diagrams, each of the terms will be at best of $O\left| \theta
\right| $, and would vanish in the limit $d\rightarrow \infty ,$ unless all
the external vertices of each daughter diagram are at the same lattice site,
that can be different for different daughter diagrams. One can conclude that
the contribution of the mother diagram would vanish when any pair of
vertices $i$ and $j$ that are not t.p.r. occupy different lattice sites,
because in that case, it is always possible to separate from the mother
diagram a daughter GF that has $i$ and $j$ in common with the rest of the
diagram. In that case, only a sum over $i=j$ should remain from the
unrestricted sum over $i$ and $j,$ and those two vertices then become t.p.r.
in the embedding of the diagram (this is illustrated in Figs. \ref{F5}b,c).
This reduction of the terms that contribute to the lattice summations has
been already described, and called ``collapse of vertices'', in the study of
the U-perturbation theory in high dimensions\cite{MHartmann89,MetznerVoll90}%
: this property will also be given the same name in the present work.

One important point to notice, is that the collapse of vertices occurs only
in the embedding of the diagram, and that the rules for calculating the
contribution should be applied to the original diagram (i.e. it would be the
diagram of Fig. \ref{F5}b in the examples given) and not to the collapsed
diagram of the embedding (i.e. the diagram in Fig. \ref{F5}c).

As in the collapse of $i$ and $j$ the two independent lattice summations
over the vertices $i$\ and $j$\ are replaced by a single lattice summation
over $i=j$, it is easy to see from the derivation of the contribution rules
in reciprocal space (cf. Section III D in Ref. \onlinecite{FFM1}) that the
two independent momentum conservation at each of the two collapsed vertices
becomes a single conservation of the wave vectors corresponding to all the
internal edges joining them. When all the pair of vertices that are not
t.p.r. have been collapsed, one can see that at every vertex of the
embedding of a vacuum or one particle GF diagram, all the edges can be
arranged in pairs with momentum ${\bf k_{s}}$ and -${\bf k_{s}}$
respectively (cf. Fig. \ref{F5}c as a typical example) It then follows that
the momentum conservation at the collapsed vertices are automatically
satisfied, and most of the usual restrains on the momentum integration
disappear in the limit $d\rightarrow \infty .$

\section{SUMMARY AND CONCLUSIONS}

\label{Sec5}The cumulant expansion of the PAM\cite{FFM1} was considered in
the limit of infinite spatial dimensions ($d\rightarrow \infty )$ for two
types of hybridization models: the purely ``local'' hybridization and the
``nearest neighbor'' (n.n.) one. As the systems usually described by the PAM
have a bandwidth much larger than the hybridization energies, the
unperturbed Hamiltonian is chosen to include the exact solution of the
conduction band electrons in the absence of hybridization. It was then
necessary to modify the derivation employed by Metzner\cite{Metzner91} for
the cumulant expansion of the Hubbard model, in which the hopping
Hamiltonian is used as perturbation. The basic result presented here is that
in spite of this change the PAM shows, for the two hybridization models
considered, the same type of simplifications that occur in the diagrammatic
cumulant expansion of the Hubbard model when $d\rightarrow \infty $.

Only the linked diagrams contribute to the general cumulant expansion, and
there is a vertex collapse when $d\rightarrow \infty $ for those diagrams
that are not f.t.p.r. A diagram is not f.t.p.r. when at least two vertices $%
i $ and $j$ can not be separated by cutting only one or two edges of the
diagram, and their collapse means that from the two independent summations
over $i$ and $j$ that are necessary to calculate their contribution, only a
summation over $i=j$ remains. The topology of the embedding is different
from that of the diagram itself when $i=j$, and the collapse is repeated
until the topology of the embedding becomes f.t.p.r. One important point to
notice is that, after the collapse of vertices, the rules for calculating
the contribution should be applied to the original diagram and not to the
collapsed diagram of the embedding.

When the two independent summations over $i$ and $j$ of the diagram become a
single summation over $i=j$ because the two vertices $i$ and $j$ collapse,
the two separate conservations of ${\bf k}$ at these vertices become a
single conservation of the ${\bf k}$ corresponding to all the edges joining
both $i$ and $j$. As a consequence, most of the usual restrains on the
momentum integration disappear in the limit $d\rightarrow \infty ,$ and the
calculation of many diagrams is very much simplified by this change. One
should note that the vertex collapse does not alter the calculation over
frequencies, which keep their conservation at all the diagram vertices, even
when $d\rightarrow \infty $.

Employing the cumulant expansion, Metzner has shown\cite{Metzner91} that the
single-particle properties of the Hubbard model in the limit $d\rightarrow
\infty $, can be described as that of independent electrons ``hopping
between dressed atoms characterized by an effective Green's function''. A
similar derivation can be employed for the PAM, and the exact one electron
GF is given by the family of diagrams in Fig. \ref{F1}a, but using an
effective cumulant $M_{2,\sigma }^{eff}(\omega )$ for the f-electron
vertices instead of the bare one $M_{2,\sigma }^{0}(\omega ).$ The effective
cumulant $M_{2,\sigma }^{eff}(\omega )$ is given by the contribution of all
the diagrams of $G_{{\sigma }}({\bf R}_{{\bf n},{\bf m}}=0,\omega )$ that
can not be separated by cutting a single edge (usually called ``irreducible
diagrams''), where $G_{{\sigma }}({\bf R}_{{\bf n},{\bf m}}=0,\omega )$ is
the exact GF of the f-electrons in the real space representation for ${\bf n}%
={\bf m.}$ This property is only valid in the limit $d\rightarrow \infty $,
so that all the diagrams that are not f.t.p.r have their vertices collapsed
until the associated embeddings are f.t.p.r. The exact GF can then be
written
\begin{equation}
G_{f,\sigma }({\bf k,}\omega {\bf )=}M_{2,\sigma }^{eff}(\omega ){\bf \ }%
\frac{1}{1-\mid V(k)\mid ^{2}G_{c{,}\sigma }^{o}(k,\omega )\ M_{2,\sigma
}^{eff}(\omega )}{\bf \qquad }  \label{E5.1}
\end{equation}

A practical difficulty in the study of correlated electron systems
with cumulant expansions, is that the higher order cumulants
rapidly become very laborious to calculate. To consider in some way
the higher order cumulants, we are studying the substitution of
$M_{2,\sigma }^{eff}(\omega )$ by an approximate quantity
$M_{2,\sigma }^{at}(\omega )$, derived from an exactly soluble
model. To this purpose we use the same Anderson periodic model but
in the atomic limit\cite{FoglioF79}, i.e. when the hopping of the
conduction electrons is eliminated by taking a conduction band of
zeroth width. This attempt will be discussed in another
publication, and the present work provides an essential guidance to
that study, by showing that all the cumulant diagrams present in
$M_{2,\sigma }^{eff}(\omega )$ are also present in the approximate
$M_{2,\sigma }^{at}(\omega )$. Although the hopping is missing from
the $M_{2,\sigma }^{at}(\omega )$, the exact solution of the
conduction band shall appear in the ``hopping between dressed
atoms''\cite {Metzner91} through the $G_{c{,}\sigma }^{o}(k,\omega
)\ $in Eq. (\ref{E5.1} ).

\acknowledgements

The authors would like to thank Prof. Roberto Luzzi for critical comments,
and to acknowledge financial support from the following  agencies:
CAPES-PICD (MSF), FAPESP and CNPq (MEF). This work was done (in part) in the
frame of the Associate Membership Programme of the International Centre for
Theoretical Physics, Trieste ITALY (MEF).

\appendix

\section{Separation of vertices}

\label{ApA}

In Section \ref{Sec4} it is shown that the contribution of a ``mother''
vacuum diagram that is not f.t.p.r. can be estimated from those of several
f.t.p.r. ``daughter'' diagram that are derived from the original one. The
daughter diagrams are obtained by separating the mother diagram into several
f.t.p.r. subdiagrams: a vacuum diagram plus several GF diagrams obtained by
adding external lines of the Bose type to the vertices that these separated
diagrams have in common in the mother diagram. To show the procedure, it is
convenient to analyze a simple case, like that of Fig. \ref{F5}a, and
consider only a single vertex in real space, as shown in Fig \ref{F6}a,
assuming first that it does not belong to the vacuum daughter diagram.
Writing only that part $P$ of the total contribution to the mother diagram
that have parameters connected to the particular vertex under consideration,
one obtains
\begin{eqnarray}
P &=&\sum_j\int d1\int d2\int d3\int d4\sum_{{\bf k_1k_2}}\sum_{{\bf k_3k_4}%
}\exp \left[ -i\left( {\bf k_1-k_2+k_3-k_4}\right) .{\bf R}_j\right]
\nonumber \\
&&\times
V(1)V^{*}(2)V(3)V^{*}(4)G_c^0(1)G_c^0(2)G_c^0(3)G_c^0(4)M_4^0(j,1,2,3 ,4)
\label{ApA.1}
\end{eqnarray}
\noindent where the $G_c^0(s)$ are the c-electron GF, and the other
abbreviations are those introduced in Section \ref{Sec4}. The time or
frequency dependence has been left out because it only plays a trivial role
in this proof. The corresponding contribution $P_a$ to one daughter diagram
for fixed values $\ell _1,\ell _2$ of the internal lines joining the vertex,
and with an added external line of momentum $K_a$ and parameters $\ell _a$
is
\begin{eqnarray}
P_a({\bf K}_a,\ell _a,\ell _1,\ell _2) &=&\frac 1{\sqrt{N}}\sum_{{\bf k_1k_2}%
}\sum_j\exp \left[ -i\left( {\bf k_1-k_2+K_a}\right) .{\bf R}_j\right]
\nonumber \\
&&\times M_3^0(j,a,1,2)\ V(1)V^{*}(2)G_c^0(1)G_c^0(2)\quad ,  \label{ApA.2}
\end{eqnarray}
and the contribution $P_b({\bf K}_b,\ell _b,\ell _3,\ell _4)$ of the other
daughter diagram is trivially obtained by replacing $a,1,2$ by $b,3,4${\sl .}

Transforming $P_{a}({\bf {K}}_{a},\ell _{a},\ell _{1},\ell _{2})$ to real
space, the contribution associated to a site ${\bf {R}}_{j}$ is given by
\begin{eqnarray}
&&P_{a}({\bf {R}}_{j},\ell _{a},\ell _{1},\ell _{2}) =\frac{1}{\sqrt{N}}%
\sum_{{\bf {K}}_{a}}\exp \left( i{\bf {K}}_{a}.{\bf {R}}_{j}\right) P_{a}(%
{\bf {K}}_{a},\ell _{a})  \nonumber \\
&&=\sum_{{\bf {k}_{1}{k}_{2}}}\exp \left[ -i\left( {\bf {k}_{1}-{k}_{2}}%
\right) .{\bf {R}}_{j}\right]
V(1)V^{*}(2)G_{c}^{0}(1)G_{c}^{0}(2)M_{3}^{0}(j,a,1,2)\quad  \label{ApA.3}
\end{eqnarray}
and a similar expression $P_{b}({\bf {R}}_{j},\ell _{b})$ is obtained for
the other daughter diagram. It is clear that the only difference between the
product $P_{a}({\bf {R}}_{j},\ell _{a},\ell _{1},\ell _{2})\ P_{b}({\bf {R}}%
_{j},\ell _{b},\ell _{3},\ell _{4})$ and the term corresponding to ${\vec{R}}%
_{j}$ and fixed $\ell _{1},\ell _{2},\ell _{3},\ell _{4}$ in Eq. \ref{ApA.1}
is given by the cumulants $M^{0}$, which are all independent of both ${\bf {R%
}}_{j}$ and of the space dimension $d$. It then follows that
\begin{eqnarray}
P &=&\int d1\int d2\int d3\int d4\ \frac{M_{4}^{0}(j,1,2,3,4)}{%
M_{3}^{0}(j,a,1,2)M_{3}^{0}(j,b,3,4)}  \nonumber \\
&&\times \sum_{j}P_{a}({\bf {R}}_{j},\ell _{a},\ell _{1},\ell _{2})\ P_{b}(%
{\bf {R}}_{j},\ell _{b},\ell _{3},\ell _{4})\quad  \label{ApA.4}
\end{eqnarray}
gives the dependence of $P$ with $d$.

When one of the two daughter diagrams is the chosen as the vacuum one, it is
necessary to consider the corresponding contribution
\begin{equation}
\overline{P}({\bf R}_j,\ell _1,\ell _2)=\sum_{{\bf k_1k_2}}\exp \left[
-i\left( {\bf k_1-k_2}\right) .{\bf R}_j\right]
V(1)V^{*}(2)G_c^0(1)G_c^0(2)M_2^0(j,1,2)\quad ,  \label{ApA.5}
\end{equation}
that was obtained replacing $M_3^0(j,a,1,2)$ by $M_2^0(j,1,2)$ in Eq. (\ref
{ApA.3}): this is the quantity that appears in the corresponding vertex of
the vacuum daughter diagram. In this case, instead of Eq. (\ref{ApA.4}) one
has the relation
\begin{eqnarray}
P &=&\int d1\int d2\int d3\int d4\frac{M_4^0(j,1,2,3,4)}{%
M_2^0(j,1,2)M_3^0(j,b,3,4)}  \nonumber \\
&&\times \sum_j\overline{P}({\bf R}_j,\ell _1,\ell _2)\ P_b({\bf R}_j,\ell
_b,\ell _3,\ell _4)\quad .  \label{ApA.6}
\end{eqnarray}

This are the relations employed in Section \ref{Sec4} to prove that only
embeddings that are f.t.p.r. give a non zero contribution, thus leading to
the collapse of vertices when they can not be separated by cutting at most
two edges.

\section{DIAGRAMS THAT ARE NOT f.t.p.r.}

\label{ApB}

In the procedure discussed in Section \ref{Sec4} a ``mother'' vacuum diagram
that is not f.t.p.r. is separated into several subdiagrams that have that
property. As discussed before, one of the daughter diagrams is chosen as a
vacuum diagram, and the remaining ones are GF diagrams obtained by adding
external lines of the Bose type to the vertices that these GF diagrams share
with other daughter diagrams in the mother diagram. Let us consider one of
the GF daughter diagrams in the reciprocal space, with momenta ${\bf {K}}%
_\upsilon ,$ indices $u_\upsilon $ and transitions $\alpha _\upsilon $ ($%
\upsilon =1,2,\ldots ,n)$ assigned to the $n\geq 2$ external vertices, where
the $\alpha _\upsilon $ correspond to pair of states with equal number of
electrons, i.e. to operators $X_\alpha $ of the Bose type. All these
external parameters will be indicated by $\{{\bf {K}}_\upsilon ,\ell
_\upsilon \}$, while $\{\ell \}_{int}$ will be used for the set of $\ell $
associated to the internal edges joining all the external vertices. The
contribution corresponding to fixed $\{\ell \}_{int}$ of a split diagram is
denoted by $F\left( \left\{ {\bf {K}}_\upsilon ,\ell _\upsilon \right\}
,\left\{ \ell \right\} _{int}\right) /N^{n/2}$, where the factor $1/\sqrt{N}$%
, associated to each external line joining an f-vertex (cf. Rule 3.7 f in
Ref. \onlinecite{FFM1}), has been explicitly written. To transform from ${%
{\bf {K}}}_\upsilon $ to position variables ${\bf {R}}_\upsilon $ one should
calculate
\begin{equation}
\overline{F}\left( \left\{ {\bf {R}}_\upsilon \right\} ,\left\{ \ell
\right\} \right) =\left( \sqrt{N}\right) ^{-n}\sum_{\{{\bf {K}}_\upsilon
\}}\left( F\left( \left\{ {\bf {K}}_\upsilon ,\ell _\upsilon \right\}
,\left\{ \ell \right\} _{int}\right) /N^{n/2}\right) \ \exp \left[ i\sum_{{%
\upsilon =1}}^n{\bf {K}}_\upsilon .{\bf {R}}_\upsilon \right]  \label{ApB.1}
\end{equation}
\noindent where $\left\{ \ell \right\} $ denotes all the $\ell $, both
internal and external. From the translational invariance of the system (or
from wave vector conservation at all the vertices) it follows that $%
\sum_\upsilon {\bf {K}}_\upsilon =0$, so that
\begin{equation}
\overline{F}\left( \left\{ {\bf {R}}_\upsilon \right\} ,\left\{ \ell
\right\} \right) =\left\{ \prod_{\upsilon =1}^{n-1}\left( \frac 1N\sum_{{\bf
{K}}_\upsilon }\exp \left[ i{\bf {K}}_\upsilon .\left( {\bf {R}}_\upsilon -%
{\bf {R}}_n\right) \right] \right) \right\} \ F\left( \left\{ {\bf {K}}%
_\upsilon ,\ell _\upsilon \right\} ,\left\{ \ell \right\} _{int}\right)
\label{ApB.2}
\end{equation}

In the two hybridization models considered in this work, the external wave
vectors ${\bf K}_\upsilon $ appear in $F\left( \left\{ {\bf K}_\upsilon
,\ell _\upsilon \right\} ,\left\{ \ell \right\} _{int}\right) $ only through
the delta that gives the wave vector conservation at each external vertex.
The internal wave vectors ${\bf k}$ appear through the $E_{{\bf k,}\sigma } $
in $G_{c,{\bf k,}\sigma }^o(\omega )$ (i.e. the c-electron GF of Eq.(\ref
{E3.9a})) and also in the $V_\sigma ({\bf k)=-\left( V_\sigma /t\right) \
E_{k,\sigma }}$ (cf. Eq. (\ref{E3.10})) for the n.n. hybridization model.
When the wave vector conservation at all vertices is considered explicitly,
the number of summations over internal ${\bf k}$ is reduced, but the
arguments of the $E_{{\bf k,\sigma }}$ become linear combinations of both
the remaining external and internal ${\bf k.}$ To illustrate this result,
consider the diagram in Fig.~\ref{F7}: applying momentum conservation one
can write
\begin{eqnarray}
&&F\left( \left\{ {\bf K}_a,\ell _b,{\bf K}_a,\ell _a\right\} ,\left\{ \ell
_1,\ell _2,\ell _3,\ell _4\right\} \right) =  \nonumber \\
&&\sum_{{\bf k_1}}M_3^0(b,1,2)\ M_3^0(a,3,4)\left| V_\sigma ({\bf k_1)}%
\right| ^2\left| V_\sigma ({\bf k_1-K_a)}\right| ^2G_{c,\sigma _1}^o({\bf %
k_1)\ G_{c,\sigma _2}^o( k_1- K_a) ,}  \label{ApB.3}
\end{eqnarray}
\noindent where the frequencies are not explicitly written because they
don't play any role in this analysis. There is only a single summation over $%
{\bf k_1,}$ because from the sum over ${\bf k_2}$ only ${\bf k_2=k_1-{K}_a}$
remains.

To study the general case, it is convenient to write (cf. Eq. (\ref{E2.4}))
\begin{equation}
E_{{\bf k,\sigma }}\equiv E_\sigma ({\bf k)=-\sum_{s=1}^d\theta _s\cos
\left( a\ k_s\right)}  \label{ApB.4}
\end{equation}
\noindent so that $F\left( \left\{ {\bf K}_\upsilon ,\ell _\upsilon \right\}
,\left\{ \ell \right\} _{int}\right) $ can be expanded in series of the $%
\theta _s$, putting $\theta _s=\theta =t/\sqrt{2d}$ at the end of the
derivation. To analyze the dependence of $\overline{F}\left( \left\{ {\bf R}%
_\upsilon \right\} ,\left\{ \ell \right\} \right) $ with $\theta $, it is
convenient to concentrate first on a given external vertex $\upsilon $, and
write ${\bf K}_\upsilon ={\bf Q}=\left( Q_1,\ldots ,Q_d\right) $ and ${\bf R}%
_\upsilon -{\bf R}_n={\bf R}=\left( R_1,\ldots ,R_d\right) .$ The Eq. (\ref
{ApB.2}) shows that for the given external vertex $\upsilon $, there is a $%
\sum_{{Q}_s}\exp \left( i\ Q_sR_s\right) $ applied to $F\left( \left\{ {\bf K%
}_\upsilon ,\ell _\upsilon \right\} ,\left\{ \ell \right\} _{int}\right) $
for each dimension $s=1,\ldots ,d$ of the space, and when for a given $s$ it
is $R_s\neq 0$, all the terms independent of $\theta _s$ in the expansion of
$F\left( \left\{ {\bf K}_\upsilon ,\ell _\upsilon \right\} ,\left\{ \ell
\right\} _{int}\right) $ will cancel out (cf. Eq. (\ref{E3.17})). It then
follows that $\overline{F}\left( \left\{ {\bf R}_\upsilon \right\} ,\left\{
\ell \right\} \right) =O\left| \theta ^p\right| $, where $p\geq p^0$ and $%
p^0 $ is the number of dimensions for which there are non zero components $s$
for at least one of the $n-1$ vectors ${\bf R}_\upsilon -{\bf R}_n$. One can
then conclude that unless all the $R_\upsilon $ coincide, the $\overline{F}%
\left( \left\{ {\bf R}_\upsilon \right\} ,\left\{ \ell \right\} \right) $is
at least of $O\left| \theta \right| $ and it vanishes when $d\rightarrow
\infty .$ This property was used to prove the collapse of vertices discussed
in Section \ref{Sec4} for diagrams that are not f.t.p.r.

\begin{figure}[p]
\caption{Typical cumulant diagrams for one-particle GF (a) The diagrams of
the chain approximation (CHA) for the f-electron, represented by the filled
square to the right, (b) As (a) but for the c-electrons , represented by an
empty square. (c) A more complicated diagram, with cumulants of fourth and
sixth order. }
\label{F1}
\end{figure}

\begin{figure}[p]
\caption{ Examples of relevant topologies in the limit $d\rightarrow \infty $%
. a)Fully two particle reducible (f.t.p.r.): any pair of vertices can be
separated by cuting two edges. Note that all pairs of loops have at most one
vertex in common. b) The diagram is not f.t.p.r. c) The topology of the
diagram and that of the embedding are the same when $i\neq j$. d) The
embedding topology that corresponds to the diagram in c) when $i=j$.}
\label{F2}
\end{figure}

\begin{figure}[p]
\caption{ a) The sum of all the simple loops of any length. b) Diagrams of
an insertion with contribution $S_2^0(j_1;\ell _1,\ell _2)$, obtained by
fixing the site $j_1$ and the parameters $\ell _1,\ell _2$ of one conduction
vertex (that is considered external) of the diagrams in a). The full circle
in these vertices, corresponding to the cumulant $M_2^0(j_1;\ell _1,\ell _2)$%
, is removed from the diagrams and is replaced by $1$ in $S_2^0(j_1;\ell
_1,\ell _2)$. c) A f.t.p.r diagram obtained by joining two simple loops at
an f-vertex. d) Family of four linked loops that give a f.t.p.r diagram. e)
Family of diagrams obtained by linking two loops at two different
f-vertices: it is not f.t.p.r. }
\label{F3}
\end{figure}

\begin{figure}[p]
\caption{ Three examples of separation of a mother vacuum diagram into
several daughter diagrams. a) Simplest case of separation with two common
vertices. b),c) Separation in three daughter diagrams with a total of four
shared vertices.}
\label{F4}
\end{figure}

\begin{figure}[p]
\caption{ a) A family of f.t.p.r diagrams that give a finite contribution
for large $d$. b)Family of diagrams obtained by joining pairs of f.t.p.r
diagrams in two common vertices; the diagrams are not f.t.p.r. c) The
f.t.p.r topology of the embedding of b) when $i=j$. Note that the diagram
contribution is calculated with the diagram b) and not with c), but
replacing the two independent summations over $i$ and $j$ by a single one
over $i=j$.}
\label{F5}
\end{figure}

\begin{figure}[p]
\caption{ Separation of a vertex with four edges into a pair of vertices
with two edges each.}
\label{F6}
\end{figure}

\begin{figure}[p]
\caption{A simple f.t.p.r diagram. The conservation of moment ${\bf k}$ at
the vertex $b$ gives ${\bf k_1-k_2-K_b=0}$ for $u_b=-1$ and that at vertex $%
a $ gives ${\bf k_2-k_1+K_a=0}$ for $u_a=+1$, so that ${\bf K}_a={\bf K}_b$
as should be expected from the translational invariance of the system. Only
one summation over ${\bf k_1}$ remains because ${\bf k_2=k_1-K_a}$. }
\label{F7}
\end{figure}

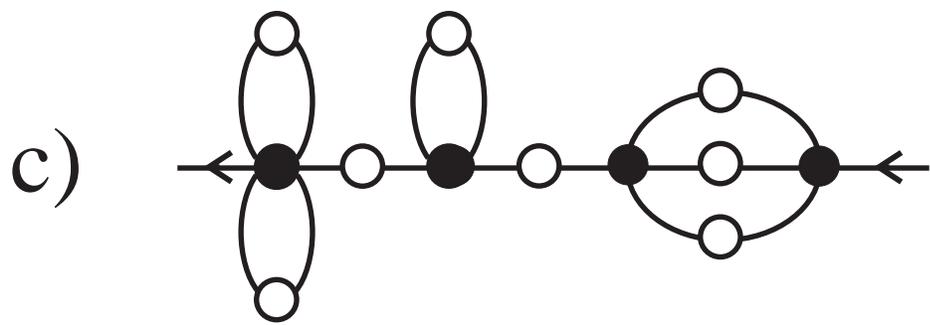
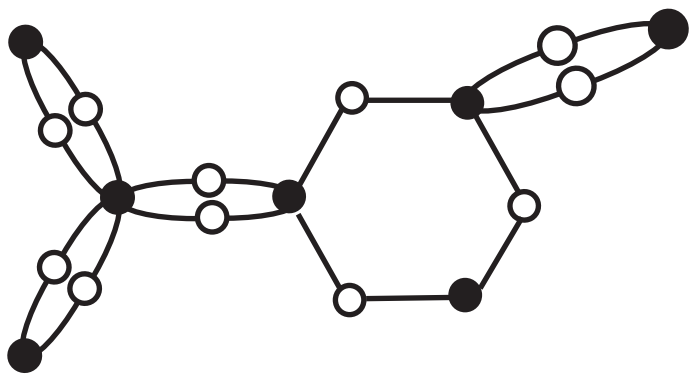
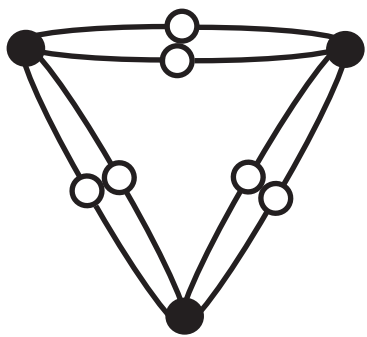
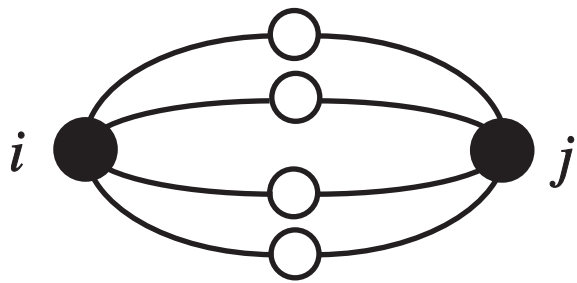
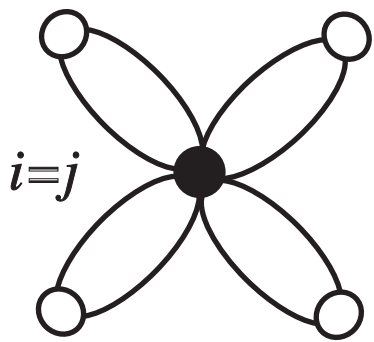
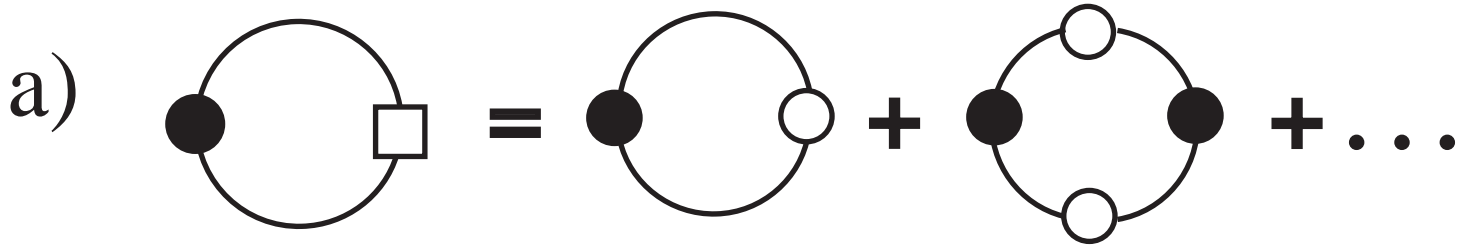
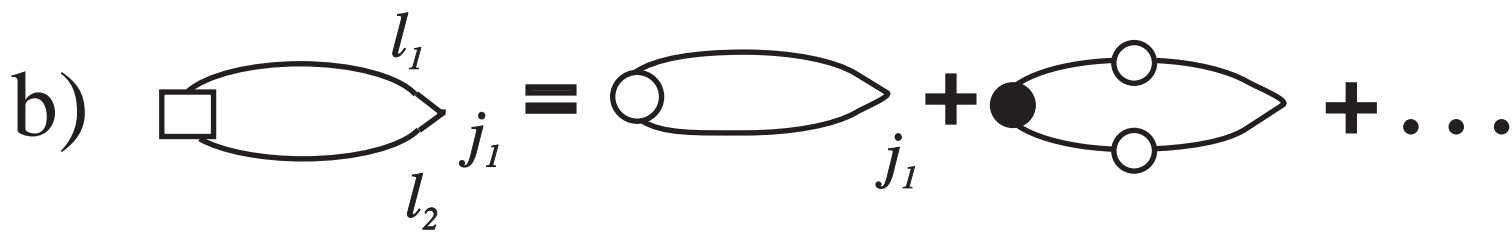
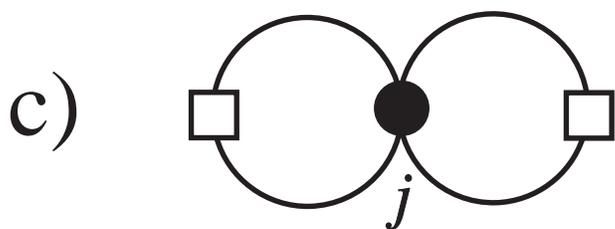
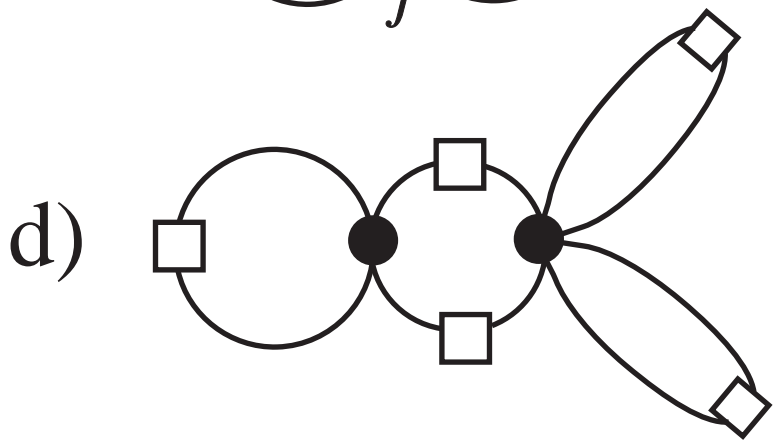
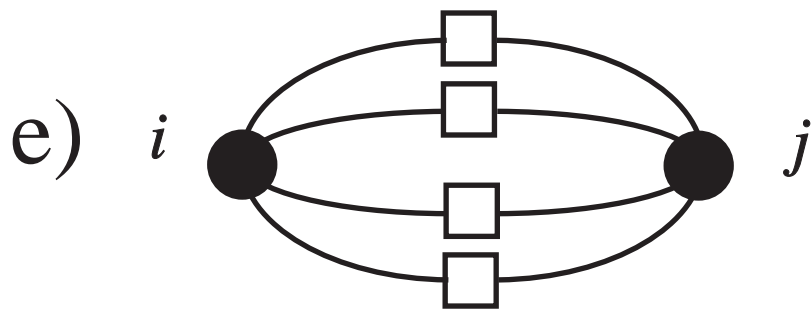
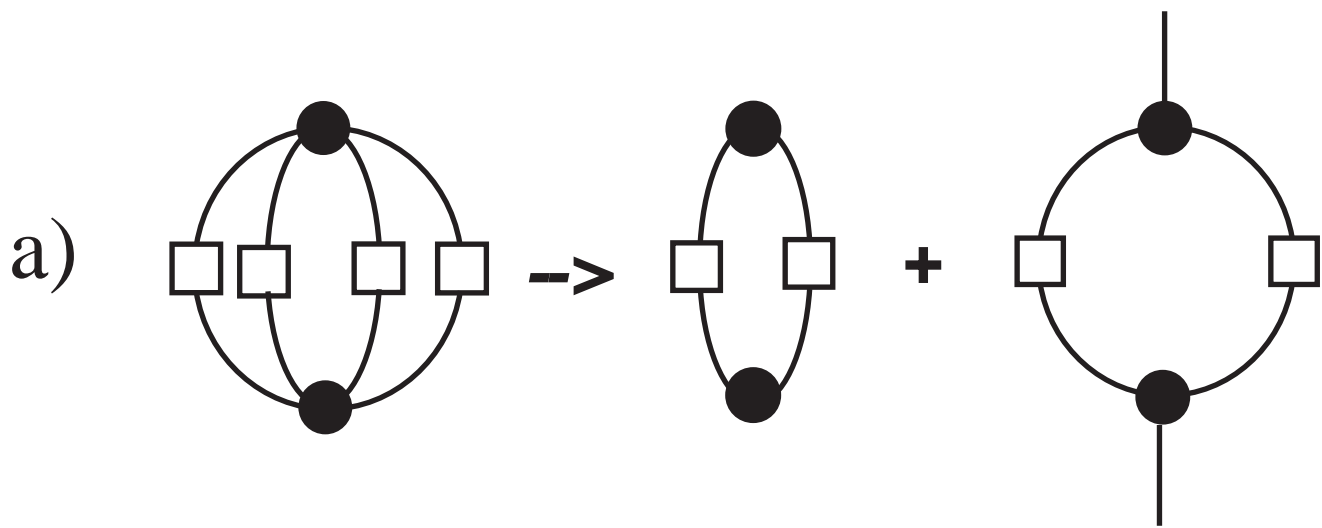
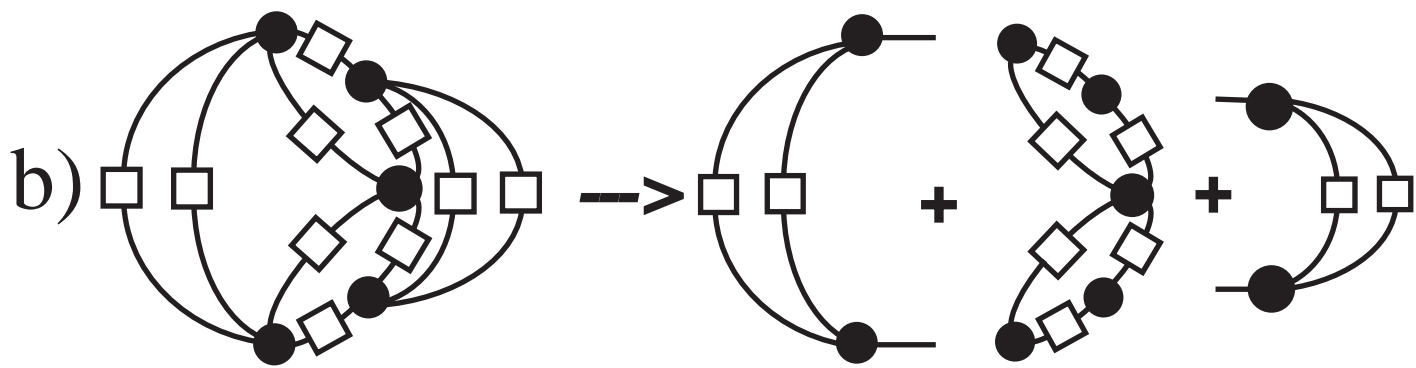
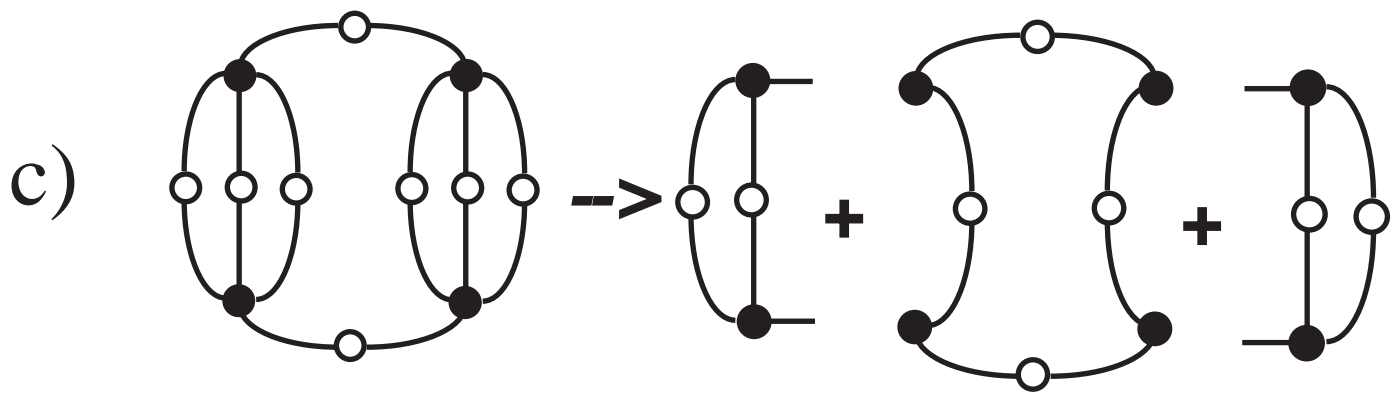
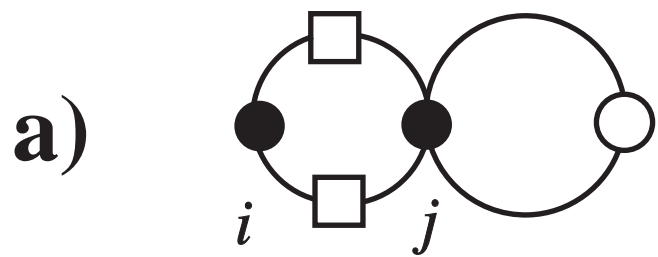
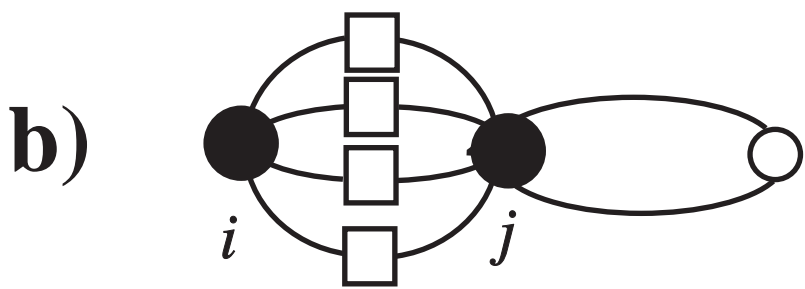
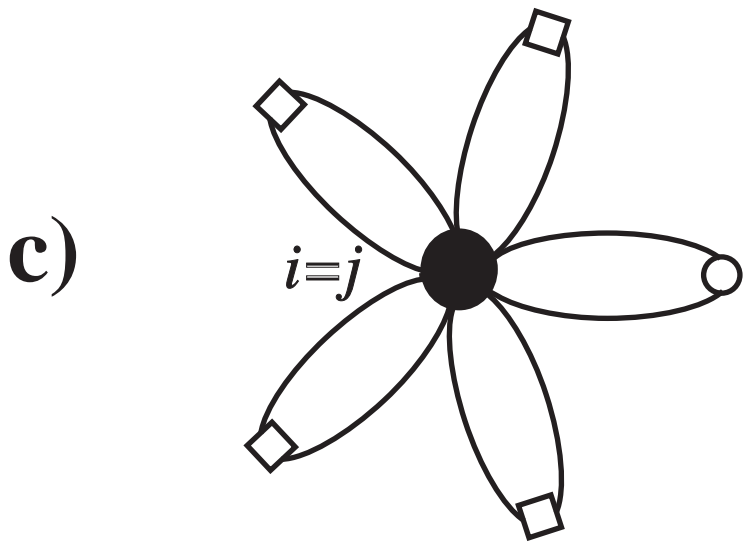
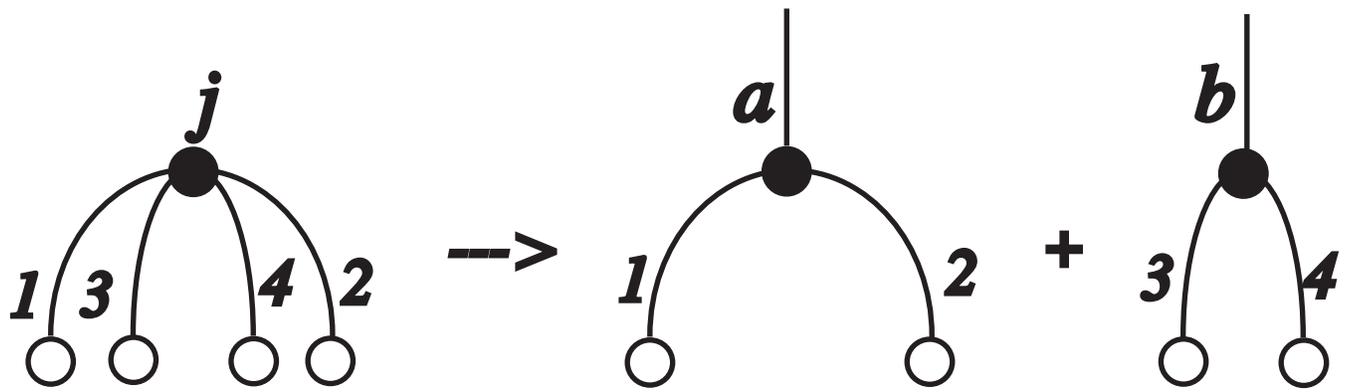
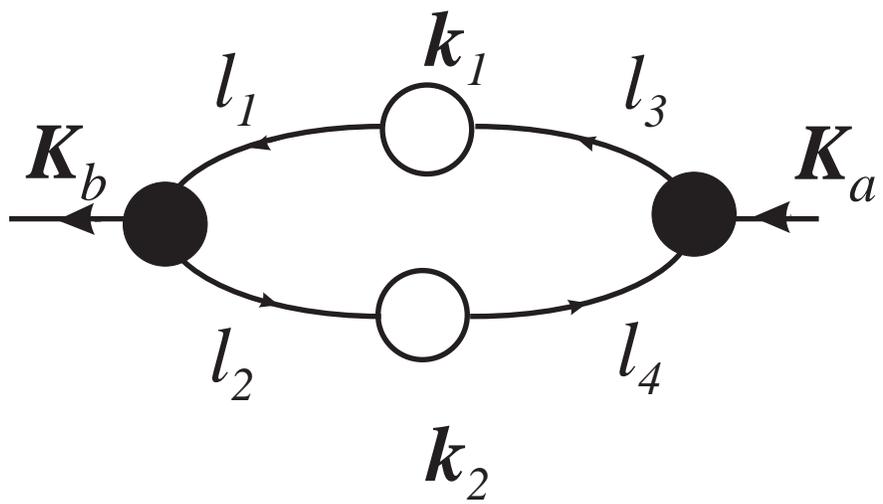
\end{document}